\documentstyle{ioplppt}                
\begin{document}
\title{ Scales of String Theory \footnote{Talk given at the Strings 99
  Conference, Potsdam, July 1999. To be published in {\em
  Class. Quantum Grav.} {\bf 17} (2000) 1 . }}
\author{C P Bachas}

\address{Laboratoire de Physique Th{\'e}orique de l' Ecole Normale
  Sup{\'e}rieure, \\ 
 24 rue Lhomond, F-75231 Paris Cedex 05, France\\
{\it bachas@physique.ens.fr}}

\begin{abstract}
I review the arguments in favor of/against the traditional hypothesis that
the Planck,  string and  compactification  scales are all within a
couple of orders of magnitude from each other. I explain how  the
extreme  brane-world scenario, with TeV type I scale and
two large (near millimetric) transverse dimensions, creates conditions
analogous  to
those of the energy desert and is thus naturally singled out. I
comment on the puzzle of gauge coupling unification in this context.
\end{abstract}

%
%

\section{Introduction}

    String/M theory \cite{GSW,pro} has a single dimensionful
parameter and a large number of dynamical moduli. The expectation
values of these moduli determine the semiclassical properties of the
vacuum, and in particular the masses of
 its Kaluza-Klein and Regge excitations. 
It is hoped that some still unknown
dynamical mechanism  ultimately selects  the (unique?) vacuum in
which we live,  thereby fixing also the scales of the new
physics. A more down-to-earth approach is to rely on arguments of
theoretical plausibility,  on direct experimental limits 
and on indirect experimental hints. In the wake of the 
`duality revolution' there has been a revival of interest on the
question of scales [3 -- 17]  which I will try to  review in the  present
talk. \footnote{The audio version and transparencies can be found in
 \cite{pro}.}

  The conventional (and conservative) hypothesis  
is that the string, compactification and Planck
scales lie  all to within two or three orders 
of magnitude from each other, and are hence far 
beyond direct experimental reach. The non-gravitational 
physics at lower energies is believed to be
described by a renormalizable supersymmetric
quantum field theory (SQFT), which  must include   the Minimal 
Supersymmetric Standard Model (MSSM). Faith in this hypothesis
has been bolstered by the following well-known 
facts~:  (i) Softly broken SQFTs
 can indeed  be extrapolated consistently to near-Planckian
energies without destabilizing the electroweak  scale~;
(ii) the minimal (or `desert') unification assumption  is in remarkable  
agreement with some of the measured low-energy  parameters of our
world, and (iii) the hypothesis is almost automatic within
the weakly-coupled heterotic string.

The negative side of the coin is that this story  is at best incomplete~:
coupling the broken SQFT to (super)gravity gives rise to vacuum
instabilities, including an unacceptably-large cosmological term. 
As has been  (re)appreciated  in recent years, our knowledge of
the gravitational interaction is in fact also very limited  in another way: 
Einstein's classical theory has
not been tested  experimentally at distances shorter than
 the macroscopic (millimeter or
$10^{-3}$ eV) regime \cite{gravi}. 
 Since this  is also the observational 
upper bound  on the  cosmological constant \cite{Wein}, it is very tempting to
speculate that a resolution of the associated
 long-standing puzzle will require a drastic modification of gravity at
such  scales. Recent proposals of a higher-dimensional
gravity \cite{ADD,RaS} 
do not  seem to point to a resolution  of
the problem \cite{BMQ},  while  more drastic modifications like  millimeter-sized  
fundamental strings \cite{mms} are  hard to accomodate in a consistent
theoretical framework.  The problem of gravitational (in)stability
is, in any case,  an important motivation for 
pursuing alternatives to the conventional  heterotic compactification scheme. 

Two other  important motivations  are
(i) that  compactification  scenaria which  confine gauge interactions
on a  brane  arise, as we have learned,  very naturally
in  controllable corners
of the  moduli space of M-theory, and  (ii) that they can bring string and
Kaluza-Klein physics closer to experiment. Furthermore, the extreme brane-world
scenario  with TeV type-I string scale \cite{Ly,AADD}
and a (near-millimetric~?)  transverse
space of dimension two, is  singled out as I will explain  in
this talk. The logarithmic sensitivity on the transverse size
\cite{B,AB} sets a stage similar to that of the conventional energy desert~:  it
makes hierarchies of scales very  natural, and 
allows reliable (model-independent)
calculations of the effective parameters in the brane theory.
 The apparent unification
of the measured low-energy gauge couplings, on the other hand,  has not found 
a  convincing explanation outside the traditional energy-desert
scenario  yet. Since this is arguably \cite{Dimo} the only clear quantitative
hint for  physics  beyond the Standard Model, 
I will start and end my 
 discussion from this point.

\section{Hints of Supersymmetric Unification}

  The well-known observation \cite{GQW,SU,LP}
  is that if one extrapolates the three
gauge couplings of $SU(3)\times SU(2)\times U(1)$ using the
$\beta$-funcions of the MSSM they meet at a scale $M_U\simeq  2\times 
10^{16}$ GeV. This is consistent with the one-loop formulae
\begin{equation}
{\alpha_i}^{-1}(\mu ) = {\alpha_U}^{-1} + b_i \; {\rm log} {\mu \over M_U}
+ \Delta_i
\label{eq:unif}
\end{equation}
where $b_i$ are the $\beta$-function coefficients, and 
$\alpha_U$ is the fine structure constant at $M_U$. 
The threshold corrections 
$\Delta_i$ parametrize  our
ignorance of  the details
of the theory at the unification scale,  and of the details of
supersymmetry breaking. 
Assuming that the  $\Delta_i$ are negligible, 
 and treating   $M_U$ and
$\alpha_U$ as input parameters, we have one
prediction which is verified by LEP data  at the level of a few
percent. 

In the minimal 
heterotic unification there is furthermore  one extra relation \cite{G}
between the  input  parameters and the experimental  value of $M_{\rm Planck}$. 
It  implies  $M_U \simeq 5\times 10^{17}$ GeV, which on the (appropriate)
logarithmic scale is a second successful prediction of the theory 
at the level again of a few percent \cite{Ka,Die}. Put
differently: there was  no a priori reason why the extrapolated low-energy
couplings should not have met, if at all, 
 at say $10^{35}$ GeV~! 
It is important here to realize that it is the very existence of the
perturbative  desert which  renders the above
  predictions meaningful and robust. 
Threshold effects make  a few-percent correction to differences 
of couplings only because  the logarithm  in equations (1)
is very large. It would be impossible  to ignore the nitty-gritty
details of the model if the unification  scale were
 say instead at $100$ TeV.

Similar arguments can be given for the mass matrices of quarks and
leptons. Minimal assumptions 
(such as discrete symmetries and Higgs-field  content) 
 determine  boundary conditions for  Yukawa
couplings, which can then be evolved with the equations of
 the renormalization group. 
The  agreement with low-energy data is  suggestive  \cite{LP},  though
less  compelling than for the gauge-coupling constants.  
Finally, I should mention that the existence of the superheavy  scale $M_U$
is indirectly supported by two other independent 
pieces of  data~: the very long lifetime of  the proton \cite{CZ} 
and the extraordinary  smallness of neutrino masses \cite{AF}.

\section{ Weakly-coupled Heterotic String}

    Let me proceed next to the theoretical arguments 
that make  the SQFT hypothesis quasi-automatic 
within the weakly-coupled heterotic string. 
Both the graviton and the  perturbative gauge  bosons
live in this case in the ten-dimensional bulk, and interact 
through the sphere diagram at tree-level. The four-dimensional  Yang-Mills and
Einstein actions therefore read
\begin{equation}
{\cal L}_{\rm gauge} \sim {(rM_h)^6\over g_h^2}\; {\rm tr} F^2 \ \ 
{\rm and}  \ \ \ 
{\cal L}_{\rm grav} \sim  {r^6 M_h^8\over g_h^2}\; {\cal R}\  ,
\end{equation}
where  $M_h$ and $g_h$ are the heterotic string  scale and string
coupling constant, while $r$ is the typical compactification 
radius. By virtue of  T-duality $r$ can be always taken 
 greater than, or equal to, the string length.
  From the coefficients of these actions  we can read the 
four-dimensional  gauge coupling and Planck mass with the result
\begin{equation}
\alpha_U \sim { g_h^2\over (rM_h)^6}\ , 
\label{eq:gin}
\end{equation}
and 
\begin{equation}
M_h^2 \sim \; \alpha_U M_{\rm Planck}^2 \ .
\label{eq:gins}
\end{equation} 
Factors of $2$'s and $\pi$'s in these relations are irrelevant for our
arguments and have been dropped. I have also dropped the
 level of the corresponding  Kac-Moody algebra which  for all practical purposes is
an integer of order one.

 Assume now  that (i) the bare gauge coupling is of order one, and
(ii)  the heterotic theory stays  weakly coupled ($g_h\leq
1$). Then  the universal  relation (\ref{eq:gins}), which we have
already encountered in the previous section,  implies that the  string
scale is tied automatically  to the Planck mass.  Relation 
(\ref{eq:gin}) on the other hand 
also  implies that $rM_h$ cannot be much larger than one.
Since it cannot be smaller than one by T-duality,  
it is necessarily of order one.
Thus  there is little  leeway for abandoning the conventional  
scenario within  the context of the weakly-coupled 
heterotic string.

 We can of course try to relax one
of the above  two assumptions~: either (i) allow $g_h$ to be
hierarchically large and hope that the gauge sector still stays under control
in  some special models \cite{A}, or (ii) 
let $\alpha_U$  be hierarchically
small   and hope that the Standard Model
gauge couplings will be  driven to their
measured values by the large threshold corrections of a higher
dimensional field theory \cite{Ba}.\footnote{Higher-dimensional
 thresholds have been
discussed also in other  contexts, see for example \cite{thr}.} 

 Such  exotic possibilities have been 
motivated in the past  by the search  for classical vacua
with broken low-energy  supersymmetry.  Known
`mechanisms' of continuous  supersymmetry
 breaking  \cite{SS,mag} indeed 
tie the breaking scale to the size of some internal dimensions.  
Classical supersymmetry restoration can   be furthermore  argued to be
singular, in string theory,  on  general grounds \cite{DS}. 
It was thus suggested early on \cite{ABLT} that one
 (or more) radii of  inverse size  at the  TeV 
would be required   if the `observed' supersymmetry breaking 
in nature were  classical.
Tree-level breaking, on the other hand, 
is at best an assumption
 of convenience -- there is no reason in principle 
why the breaking in nature should not
have a non-perturbative origin. 
Furthermore the classical mechanisms  
have  not  so far lead to  new insights on the crucial problems of vacuum 
selection and stability. Thus, there seemed  to be little theoretical
motivation for abandoning the conventional 
compactification scheme,  and
its successful unification predictions,  
in  heterotic string theory.


\section{Brane World and Type-I  Theory}

   One of the important developments
of the `duality revolution' has been the 
realization that various branes -- Dirichlet branes \cite{Po,D}, or their dual
heterotic fivebranes \cite{Wz} and  Horava-Witten walls \cite{HW} --
can trap non-abelian gauge interactions in their
worldvolumes. This has placed on  a firmer basis  an old idea
 \cite{RS} according to which 
 we might be  living  on a  brane 
embedded in a  higher-dimensional world. 
The idea arises  naturally  in
compactifications of type I theory \cite{DLP},  which typically involve
collections of orientifold planes and D-branes. 
I will from now on restrict my discussion to this context, 
because in it the `brane-world' scenario admits a fully
perturbative string description. All   other interesting
possibilities (see for example  \cite{RaS,AP,BO})  
involve some type of non-perturbative dynamics and are a priori harder
to control.

  In type I string theory  the graviton (a closed-string state) 
lives  in the ten-dimensional bulk, while open-string vector bosons 
are in general  localized on  lower-dimensional D-branes. Furthermore 
while closed strings interact
to leading order via the sphere diagram,  open strings 
 interact via  the disk diagram which is of higher
order in the genus expansion.
The four-dimensional Planck mass and Yang-Mills couplings
therefore take the form
 \begin{equation}
\alpha_U \; \sim \; {g_{\rm I}\over ({\tilde r} M_{\rm I})^{6-n}}  \ \  , \ \ \ 
M_{\rm Planck}^2 \sim { r^{n}{\tilde r}^{\; 6-n} M_{\rm I}^8\over g_{\rm I}^2} ,
\label{eq:typei}
\end{equation}
where $r$ is the typical radius
 of the $n$  compact dimensions transverse to the 
brane,   $\tilde r$ the typical radius
 of the remaining (6-$n$) compact longitudinal
dimensions, $M_{\rm I}$ the type-I string scale and $g_{\rm I}$
the string coupling constant. By appropriate T-dualities we can again
 ensure  that both $r$ and $\tilde r$ are
greater than or equal to the fundamental  string scale. T-dualities
 change $n$ and  may take us either to
Ia  or to Ib theory (also called I or I',  respectively) but  I will
not make a distinction between these two.

It follows from these formulae that
  (i) there is no universal relation between
$M_{\rm Planck}$, $\alpha_U$ and $M_{\rm I}$ anymore, and (ii) tree-level
gauge couplings corresponding to different sets
of D-branes have radius-dependent  ratios and 
need not  unify at all.    
Thus  type-I string theory is
much more flexible (and  less predictive)  than its
 heterotic counterpart.
The fundamental string scale,  $M_{\rm I}$,  in particular  is 
 a free parameter, even if one insists  that
 $\alpha_U$  be  kept fixed and
of order one,  and that the string theory be weakly coupled.
This added flexibility can be used to `remove'  the
order-of magnitude discrepancy
between the apparent unification and string scales of the heterotic theory
\cite{W}, 
to lower $M_{\rm I}$ to an intemediate scale  \cite{Kar,BIQ} or
even all the way down to its 
experimentally-allowed limit of order the TeV  \cite{Ly,AADD}.
Keeping for instance $g_{\rm I}$,
$\alpha_U$ and $({\tilde r} M_{\rm I})$ fixed and of order one, 
leads to the condition
\begin{equation}
r^n \sim {M_{\rm Planck}^2/ M_{\rm I}^{2+n}} .
\label{eq:mm}
\end{equation}
A TeV string scale would then require from  $n=2$  millimetric 
to  $n=6$  fermi-size dimensions transverse to our brane world. The
relative weakness of gravity is  in this picture 
attributed to  the transverse spreading of gravitational flux.

\section{Experimental Bounds}

  What has brought the brane-world idea 
 into focus \cite{ADD} was the realization
that it cannot be a priori ruled out by the existing data,  even in
 the most  extreme case of `TeV-ish'  string scale and 
millimmeter-size  transverse dimensions. 
Gravity is hard to test  at submillimeter  distances because of the
 large  background of residual
 electromagnetic interactions. The ratio for instance of the  Van der
 Waals to Newtonian force 
 between two hydrogen atoms a distance $d$ apart is \cite{ADD}
\begin{equation}
{F_{\rm VdW}\over F_{\rm grav}} \sim 
\left( {1 { mm}\over d} \right)^5 \ .
\end{equation}
At    $d= 10\mu m$ Newton's force is thus ten orders of magnitude
weaker than Van der Waals~!  
As a result the present-day data  \cite{gravi} allows  practically any
modification of Newton's law,  as long as it is  of
comparable strength at and screened beyond the millimeter range. 
This has been 
 appreciated in the past in the context of gravitational
axions \cite{axi}, and similar bounds hold for light  string moduli \cite{mo} 
or  extra
 Kaluza-Klein dimensions \cite{CKM,ADD,KSf}.

       Besides mesoscopic gravity experiments, there are two other 
types of direct experimental limits 
one should worry about~: 
those coming from  precision observables  of the
 Standard Model, and those coming from various  exotic processes. 
Precision  tests of the SM and
compositeness bounds cannot  rule out in a model-independent way any
new physics above the TeV-ish  scale. 
Bounds for instance from LEP data on
four-fermion operators, or bounds on dimension-five operators
contributing to the  $g-2$ 
of the electron/muon are  safe, as long as  the characteristic scale of
the new physics is a few TeV \cite{ADD}. 
Proton decay and other exotic processes could  of course
rule out large classes of low-scale models. There exist however plausible
 suppression mechanisms,  such as  bulk U(1) gauge symmetries  which are 
 spontaneously-broken at some distant brane \cite{ST,proton} and
look like approximate global symmetries  in our brane world. 
One type  of model-independent 
exotic process is  graviton emission  in the bulk, which could be seen
as missing-energy events in  collider experiments 
\cite{missing}. The process  is however
suppressed by the four-dimensional Newton constant  at low
 energies,  and only becomes
 appreciable (as one should expect) 
near string  scale  where quantum gravity effects
 are strong. 

None of these (or other) phenomenological considerations  seems a
 priori fatal to the brane-world scenario, even in its  extreme
 realization. Put together they will, however, probably  make  
 realistic type-I model building  a
 very strenuous  exercise indeed.

\vfil\eject

\section{The Trasverse Desert}

  Although  $M_{\rm I}$ could lie anywhere between the Planck mass and
  the  TeV , lowering it to the latter scale has two advantages~: (i)
  it brings string physics
 within the reach of future acceleretor experiments, and (ii) it is 
a natural starting point for
 discussing the problem  of the gauge hierarchy, which becomes now a
  question in the infrared \cite{ADD,AB}.  In a certain
 sense  this extreme choice is antipodal to the energy-desert scenario:
although the MSSM is a stable  renormalizable field theory, we are shrinking 
its range of validity  to  one order of magnitude at most!
 Nevertheless, as I will now argue, these two scenaria
 share many common features  when the number of large
 dimensions transverse to our brane is exactly  two \cite{B,AB}.

   The key feature of the SQFT  hypothesis is that low-energy
   parameters receive large logarithmic corrections, which are
  effectively resummed  by the equations of the Renormalization Group. 
  The logarithmic sensitivity of parameters
  also generates naturally  hierarchies of
 scales, and has been  the key ingredient in all
  efforts to understand the origin of 
 the gauge hierarchy in the past \cite{N}. 
Consider now  the brane world scenario. The parameters of the
 effective brane lagrangian are dynamical open- and closed-string
 moduli with constant expectation values along the four non-compact
 space-time dimensions of our world. The closed-string 
moduli, $m_a$,  are bulk fields
 whose expectation values will generically vary as a function of the
 transverse coordinates $\xi$. They include the dilaton,
 twisted-sector massless scalars, the metric of the transverse space etc.
For weak  type-I string coupling these variations can be 
 described by a lagrangian of the form
 \begin{equation}
{\cal L}_{\rm bulk} + {\cal L}_{\rm source}
\sim \int d^n\xi \; \Bigl[ {1\over g_{\rm I}^2}
(\partial_\xi  m_a)^2 +  {1\over g_{\rm I}} \sum_s f_s(m_a) 
\delta(\xi-\xi_s)\Bigr]
.
\label{eq:bb}
\end{equation}
Here ${\cal L}_{\rm bulk}$ is a reduced supergravity Lagrangian 
while the sources are the  D-branes and orientifolds which are  localized
at  positions $\xi_s$ in the transverse space. The couplings
$f_s(m_a)$ may vary  from source to source -- they can for instance
depend on open-string moduli --  and are subject to global
consistency  conditions. What is  important is that they are
 weak  in the type-I limit, leading to weak field variations, 
 \begin{equation}
m_a(\xi) = m_a^0 + g_{\rm I}\;  m_a^1(\xi) + \cdots  , 
\label{eq:bb}
\end{equation} 
with  $m_a^0$  the (constant) 
 average value, $m_a^1(\xi)$  given by a sum of Green's functions, and
so on.  For $n=2$  transverse dimensions the leading variation
 grows  logarithmically with the size,  $r$, 
 of  the transverse space.  Since our
Standard Model parameters will be  a function of the moduli evaluated at
the position of our brane world, they will have  logarithmic
sensitivity on $M_{\rm Planck}$, very much like  the (relevant)
parameters of  a  supersymmetric renormalizable QFT. Similar
sensitivity may  occur even if $n>2$ ,  as long as  some of the `bulk'
moduli propagate  in only two extra large dimensions.

The bulk  supergravity Lagrangian  receives 
both stringy and higher-genus corrections, but these involve 
higher derivatives of fields, and should therefore be 
negligible for moduli  varying logarithmically over distances
much larger than the string scale.
 The source functions, $f_s(m_a)$, will also
be  generically  modified by such corrections -- the D-branes have
indeed  string-scale thickness when probed by the supergravity fields
\cite{probe}. Such  source modifications can, however, be absorbed
into boundary conditions for the  supergravity equations,    
at the special marked points $\xi_s$. The situation thus looks (at
 least superficially) analogous to that prevailing under 
 the SQFT hypothesis~: large  corrections to effective low-energy couplings
 can be in both cases  resummed by  differential equations
 subject to appropriate boundary conditions. 
Furthermore `threshold corrections'
parametrizing our ignorance of the detailed physics at distant branes
-- the analog of physics near the unification scale -- 
have a small effect on the relative evolution of parameters,
provided  the transverse two-dimensional space is
sufficiently large. Clearly $n=2$ is
critical: for exactly one  transverse dimension bulk fields 
vary linearly in space and one expects to hit strong-coupling
singularities before $r$ can  grow very large,
while for $n>2$  the 
dynamics on our  brane  completely decouples from the dynamics elsewhere
in the bulk.

\section{The Puzzle of Unification}

 The logarithmic sensitivity  of brane parameters with
$r$ seems a natural setting for   generating  scale hierarchies dynamically,
as in the case of  renormalizable SQFT.  Gauge dynamics   on
a  given brane, for example, could  become strong as  the transverse
space  expands to an exponentially  large size, thereby inducing  gaugino
condensation and  supersymmetry breaking. The problems of vacuum
selection and stability are, to be sure, still with us -- the
situation is, in this respect,  neither better nor worse than 
in the traditional compactification scenario. 

 The apparent unification of the (MS)SM  gauge  couplings, on the
other hand, has not yet found a convincing `explanation' in this (or
any alternative) context. The basic  ingredients are, nevertheless,
present \cite{B} (see also \cite{I,ADM} for related ideas): (i) the
logarithmic variation of bulk fields in the large transverse space can
give equally-robust predictions for brane parameters, as
renormalization group  running
over the energy desert, (ii) some bulk fields (like twisted scalar
moduli) have non-universal couplings to gauge fields that live on the
same set of D-branes \cite{AFIV} -- they 
 can thus split their gauge couplings apart
without separating them in transverse space, and (iii) the boundary
condition imposing  unification  at high energy could be replaced
by a boundary  condition that these non-universally coupled bulk moduli
vanish on some far-distant brane (for instance because of a
non-perturbative superpotential). The main difficulty of  these ideas
is to explain   why 
the coefficients of the logarithmic real-space evolutions  should
have  the same (ratio of?) differences as the MSSM beta
functions. This is clearly required if we are to `understand' the actual
 values of the  SM gauge couplings, as measured at LEP energies.

\vspace{.75cm}

{\bf Aknowledgements}

 I thank the organizers for the invitation to speak, and for
organizing a marvellous strings 99  conference.
 This work was partially supported
by the TMR contract ERBFMRX-CT96-0090.

\vspace{.75cm}

\vfil\eject

\end{document}